# Comprehensive landscape and simple rules for transition-metal Heusler semiconductors

Yubo Zhang[1,*], Da Ke[1], Zirui Dong[2,*], and Jun Luo[3,*]

[1]*Minjiang Collaborative Center for Theoretical Physics, College of Physics and Electronic Information Engineering, Minjiang University, Fuzhou 350108, China*

[2]*School of Materials Science and Engineering, Shanghai University, Shanghai 200444, China*

[3]*Interdisciplinary Materials Research Center, School of Materials Science and Engineering, Tongji University, Shanghai 201804, China*

**Corresponding email: yubo.drzhang@mju.edu.cn, ziruidong@shu.edu.cn, junluo@tongji.edu.cn*

Heusler alloys, renowned for their multifunctionality and capacity for vast elemental customization, are primarily classified into half-Heusler (XYZ) and full-Heusler ($X_2YZ$) structural types. Typically, the 18-electron half-Heusler and the 24-electron full-Heusler alloys are recognized as semiconductors, following the Slater-Pauling rule. Semiconductors are desired for many applications, but they represent a minor portion compared to the predominantly metallic and half-metallic members of the Heusler family. Recently, *vacancy-filling off-stoichiometric Heuslers* of ternary $X_{1+\beta}YZ$ ($0 \le \beta \le 1$) and quaternary $X_\alpha X'_\beta YZ$ ($1 \le \alpha + \beta \le 2$) have emerged as a versatile strategy to broaden the scope of Heusler semiconductors. However, the flexibility associated with off-stoichiometry inevitably leads to complications, including issues with fractional filling ratios and complex site occupations. This work presents a comprehensive landscape of transition-metal-containing Heusler semiconductors, focusing on the off-stoichiometric Heuslers but seamlessly encompassing the integer-stoichiometric systems. The structural and electronic properties can be theoretically understood through a few simple rules. Many systems have been experimentally validated, showcasing their potential for applications such as thermoelectric converters.

## 1. Introduction

Heusler alloys [1] are traditionally categorized into two main structural types for ternary systems, i.e., half-Heusler (XYZ) and full-Heusler ($X_2YZ$), with integer stoichiometries of 1:1:1 and 2:1:1, respectively. In transition-metal Heuslers, following the nomenclature by pioneering work of Galanakis [2,3,4], X represents high-valent elements (e.g., Fe, Co, and Ni) and is more electronegative, Y is a low-valent transition-metal element (e.g., Ti and V) and is more electropositive, while Z represents a main-group element with *sp* valence electrons (e.g., Al, Si, and Sb). The half-Heusler structure comprises three interpenetrating face-centered-cubic sublattices occupied by X, Y, and Z elements, whereas the full-Heusler features a fourth sublattice occupied by an additional X. In half-Heuslers (Figure 1a), the Z and Y elements occupy respectively the 4*a* and 4*d* Wyckoff positions, forming a rock-salt framework through ionic-like bonds. The electronegative X elements enter the 4*c* interstitial positions, having strong covalent interactions with Y and Z elements, their nearest neighbors. In $X_2YZ$ full-Heuslers (Figure 1a), both the 4*c* and 4*d* sites are filled by X, leading to a symmetrized structure at the 8c sites. The X-X chemical hybridization, although involving the second-nearest neighbors, is crucial for understanding the magnetic and electronic properties of the full-Heuslers [3]. The quaternary variant XX′YZ, in which X and X′ represent two inequivalent elements, adheres to the integer stoichiometry of 1:1:1:1.

Thousands of Heusler alloys can be obtained through flexible elemental substitution, and their fundamental magnetic and electrical properties generally adhere to the well-known Slater-Pauling rule [5,6]. Typically, half-Heuslers with $N_t$ = 18 valence electrons (per formula unit) are nonmagnetic semiconductors (e.g., CoTiSb and NiTiSn), while other electron counts lead to ferromagnetic half-metals with a net magnetization of $M_t = N_t - 18$ [2]. The corresponding rule for full-Heuslers is $M_t = N_t - 24$ [3], where 24 electrons result in semiconductors (e.g., $Fe_2VAl$). This rule has been generalized to quaternary full Heuslers, such as *inverse Heuslers* [4] and *LiMgPdSn-type Heuslers* [7]. Assisted by the Slater-Pauling rule, researchers have identified some exotic materials beyond conventional nonmagnetic semiconductors, including ferromagnetic semiconductors, fully-spin-compensated ferrimagnetic semiconductors, spin-gapless semiconductors, half-metallic antiferromagnets, and topological insulators [8,9,10].

Semiconducting Heuslers, though critical for various applications like thermoelectrics, represent only a tiny fraction of the predominantly metallic and half-metallic Heusler family. Many efforts have been made to transform the 17- and 19-electron metallic half-Heuslers into 18-electron semiconductors. Anand et al. have theoretically proposed the *double-half-Heusler* concept that adheres to the conventional 18-electron rule by combining a 17-electron FeTiSb with a 19-electron NiTiSb, leading to an *averaged* material of $FeNiTi_2Sb_2$ or $Fe_{0.5}Ni_{0.5}TiSb$ [11,12]. In this material, the high-valent Fe and Ni elements mix and occupy the 4*c* sites, allowing $Fe_{0.5}Ni_{0.5}TiSb$ to conform to the half-Heusler structure. Similarly, among the low-valent elements, occupation mixing can occur at the 4b sites. An example is $NiMg_{1-x}Ti_xSb$, whose endpoints are the 17-electron NiMgSb and 19-electron NiTiSb. The intermediate composition of $NiMg_{0.5}Ti_{0.5}Sb$ is a semiconductor with 18 electrons [13]. For the 19-electron half-Heusler, a more straightforward approach involves reducing one electron through vacancies: Zhu et al. introduced 20% vacancies onto Nb's sites (4*b* sites) in CoNbSb, resulting in the 18-electron $CoNb_{0.8}Sb$ [14,15]. Similarly, the 17-electron half-Heusler can absorb an additional electron to achieve the 18-electron configuration. Wolverton et al. utilized lithium as the electron donor, and they theoretically realized quaternary Heuslers belonging to the LiMgPdSn-type [16]. Other quaternary semiconductors with either 18 or 24 valence electrons have been recently identified from brute-force screening [17]. Note that the first three example materials (i.e., $Fe_{0.5}Ni_{0.5}TiSb$, $NiMg_{0.5}Ti_{0.5}Sb$, $CoNb_{0.8}Sb$) still belong to the half-Heusler category despite their non-stoichiometric appearance, which is distinct from a new type of Heuslers to be discussed later.

*Vacancy-filling off-stoichiometric Heuslers* (Figure 1a), which eschew the constraints of integer stoichiometry, have emerged as a more versatile strategy for realizing semiconducting behaviors. Instead of compromising the integrity of the rock-salt framework, this approach focuses on engineering the filling of interstitial vacancy sites, aiming for an occupation ratio intermediate between half- and full-Heuslers. For example, ternary $X_{1+\beta}YZ$ (with $0 < \beta < 1$) represents the fractional vacancy filling ratio, and the half- and full-Heuslers are recovered at the endpoints (i.e., $\beta = 0$ and $\beta = 1$). $Fe_{1.5}TiSb$ was the first system theoretically predicted in 2016 [18], although experiments revealed poor-quality samples that exhibited significant defects and secondary metallic phases [18,19]. The bandgap mechanism of $Fe_{1.5}TiSb$ was later detailed by Snyder and colleagues [20], a topic we will revisit in this work.

Our team has worked extensively on these materials in recent years. For instance, we synthesized a similar material, $Ru_{1.5}ZrSb$, which exhibits clear semiconductor behaviors [21]. We also found that $Ru_{1+x}TiSb$ displays semiconductor or near-semiconductor properties across various compositions [22]. Additionally, we designed and realized the quaternary $Fe_xCu_yTiSb$ [19,25], which transits from good to poor semiconductors by varying the composition [23]. We observed exotic behaviors such as non-Fermi-liquid behavior in $Fe_{1.33}TiSb$ [19] and Kondo-like magnetic phenomena in Cr-doped $Fe(Nb_{0.75}Ti_{0.25})Sb$ [24], both deviating slightly from the optimal semiconducting stoichiometries.

While the *vacancy-filling* strategy significantly expands the range of Heusler semiconductors, it also poses substantial challenges in understanding the materials associated with the off-stoichiometry. Experimentally, the systems that have been explored represent only a small fraction of potential candidates [18,19,21,22,23,24], suggesting that numerous others await investigation. Previous theoretical research has primarily concentrated on already synthesized systems, aiding in interpreting experimental observations [19,20,23,24]. The incomplete theoretical understanding limits its applicability in guiding new practitioners effectively. In this context, the current work presents a comprehensive landscape of Heusler semiconductors, focusing on systems containing transition-metal elements. As we explore the off-stoichiometric formulations, the resulting landscape naturally includes the integer-stoichiometric formulations as specific cases. Simple rules are proposed to understand the structural and electronic properties, and some selected materials are experimentally validated. Our findings may inspire further exploration and development of these materials.

## 2. Simplified understanding of the bandgap mechanisms

Before addressing the bandgap mechanism, we categorize the Heusler compounds based on their composition and stoichiometry. As depicted in Figure 1b, the general formula can be expressed as $X_{\alpha}^{4c}X'^{4d}_{\beta}Y^{4b}Z^{4a}$ (X and X′ are high-valent transition-metal elements), with the 4*c*-4*d* interstitial occupations following a constrain of $1 \leq \alpha + \beta \leq 2$. The familiar integer-stoichiometric Heuslers include XYZ for the *ternary half-Heusler* (THH) and $X_2YZ$ for the *ternary full-Heusler* (TFH). *Quaternary half-Heusler* (QHH) is expressed as $X_{\alpha}X'_{1-\alpha}YZ$, including the intermediate $X_{0.5}X'_{0.5}YZ$ for the double-half Heusler [11]. *Quaternary full-Heusler* (QFH) is expressed as $X_{\alpha}X'_{2-\alpha}YZ$, where XX'YZ represents the well-studied quaternary Heusler. We use $X_{1+\beta}YZ$ to represent the *ternary off-stoichiometric Heusler* (TOSH) and $X_{\alpha}X'_{\beta}YZ$ for the *quaternary off-stoichiometric Heusler* (QOSH). One objective of this work is to determine the filling coefficients that lead to a semiconducting phase, characterized by specific electron counts and atomic occupation patterns.

Heusler's bandgap is intricately linked to the formation of fully occupied hybrid orbitals through covalent bonding interaction, and the mechanisms in THH and TFH have been detailed using orbital hybridization diagrams [2,3]. Figure 1c schematically shows the $s^2p^6d^{10}$ hybrid orbitals for THH $X^{4c}Y^{4b}Z^{4a}$, where the bandgap is above the occupied orbitals. When XYZ has fewer than 18 electrons ($N_t < 18$, see Figure 1f), the spin-down *d*-shells are incomplete, resulting in half-metallicity and net magnetization. To compensate for the electron deficiency in XYZ, the *vacancy-filling strategy* introduces a fraction of X transition-metal atoms into the 4*d* interstitial sites, resulting in the chemical formula $X^{4c}Y^{4b}Z^{4a}$+$X_{\beta}^{4d}$ or equivalently $X_{1+\beta}YZ$ ($0 \leq \beta \leq 1$). The transition from Figure 1f to 1c, which is conceptually straightforward and has been utilized in lithium-doped systems [16], is unrealistic for the vacancy-filling approach based on transition-metal elements. The reason is that the transition-metal ion at the 4*d* sites—typically high-valent transition metals—tends to retain some *d*-electrons as valence electrons; but, low-valent transition metals are usually too electropositive to occupy interstitial sites [9]. These 4*d*-site elements, after donating electrons to the XYZ framework, must also create *local gaps* to render the entire structure insulating. The gapping mechanism is termed as $s^2p^6d^{10}$+$\beta\times d^6$ (Figure 1d): the inserted 4*d*-site ion experiences Coulombic repulsion from six neighboring 4*c* ions (Figure 1h), leading to octahedral crystal-field splitting between $t_{2g}$ and $e_g$ orbitals. Although the 4*c* and 4*d* sites are next-neighboring, their *d*-orbital interaction can be significant due to the directional nature of the $e_g$ orbitals [3]. This mechanism applies to the TOSH $Fe_{1.5}TiSb$ [18,20] and $Ru_{1.5}ZrSb$ [21].

When XYZ has more than 18 electrons ($N_t > 18$, see Figure 1g), three mechanisms can potentially open a bandgap. The first approach involves reverting XYZ to the 18-electron configuration by removing excess electrons (from Figure 1g to 1c). The aforementioned $CoNb_{0.8}Sb$ belongs to this category [14,15]. The second approach is to transform Figure 1g to 1d via vacancy filling, but it is practically challenging due to the unique electronic characteristics of the transition-metal elements, as detailed in the Supplementary Materials (**Section 8**). A more promising approach is termed $s^2p^6d^{10}$+$\beta\times d^{10}$ mechanism (Figure 1e) associated with vacancy filling. In this scenario, the doped 4*d* atoms initially absorb excess electrons from the $X^{4c}Y^{4b}Z^{4a}$ matrix, leaving 18 electrons; subsequently, the resulting 4*d* ions can develop *local gaps* if their *d*-shells become fully filled. Practically, the doped X′ element must be different from the X element in the host, leading to QOSH $X^{4c}Y^{4b}Z^{4a}$+$X'^{4d}_{\beta}$.

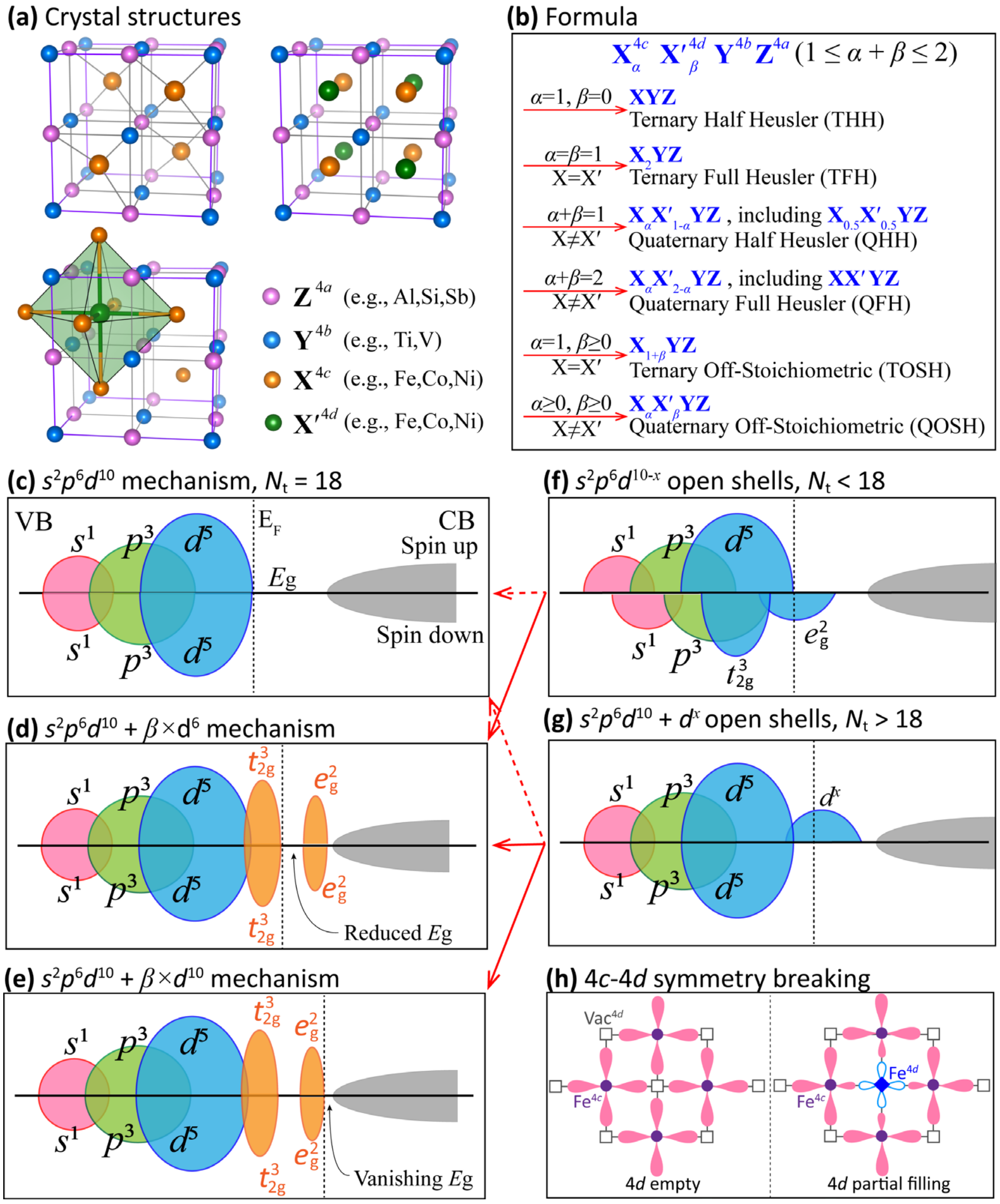

**Figure 1. Simplified gapping scenarios in transition-metal Heusler semiconductors.** (a) Crystal structures of half-, full-, and vacancy-filling off-stoichiometric Heuslers. In the last structure, an octahedron is plotted to highlight the interaction between 4*c* and 4*d* sites. (b) Categorization of Heusler materials. (c) A cartoon plot illustrates the electronic density-of-states and the $s^2p^6d^{10}$ bandgap mechanism of 18-electron semiconductors. (d) Bandgap formation due to $t_{2g}$-$e_g$ octahedral field splitting, leading to an $s^2p^6d^{10}$+$\beta\times d^6$ mechanism. (e) Bandgap mechanism of $s^2p^6d^{10}$+$\beta\times d^{10}$. (f, g) Metallic Heuslers with open *d*-shells. (h) Breaking of 4*c*-4*d* geometrical symmetry in off-stoichiometric Heuslers because the 4*d*-site orbitals experience more next-neighboring repulsions than the 4*c*-site orbitals. Solid lines between subplots denote bandgap strategies via vacancy-filling, whereas dashed lines illustrate alternative strategies.

While the $s^2p^6d^{10}$+$\beta\times d^{10}$ mechanism is clear and straightforward, the $s^2p^6d^{10}$+$\beta\times d^6$ mechanism warrants further explanation, as exemplified in $Fe_{1.5}TiSb$ (Figure 2). We have interpreted the $s^2p^6d^{10}$+$\beta\times d^6$ mechanism as "full-$t_{2g}$ and empty-$e_g$" orbital configurations. Applying this interpretation to $Fe_{1.5}TiSb$ would suggest nominal orbital configurations of $Fe^{4c}$-$d^{10}$ and $Fe^{4d}$-$d^6$, indicating the presence of distinct species of $(Fe^{4c})^{2+}$ and $(Fe^{4d})^{2-}$ [20]. However, the calculated electronic density of states (Figure 2b) reveals stark contradictions. First, the integrated electron numbers for $Fe^{4c}$ and $Fe^{4d}$ ions are nearly identical, at 7.13 and 7.02 electrons, respectively. Second, the $Fe^{4d}$-$e_g$ orbitals retain a considerable number of electrons (approximately 1.93 electrons per atom) instead of presenting an empty configuration. These inconsistencies imply that the interpretation of "$t_{2g}$-$e_g$ splitting in an octahedral field" (Figure 1a) is an oversimplification for $Fe_{1.5}TiSb$, though conceptually simple and useful.

In $Fe_{1.5}TiSb$, the atomic interactions include nearest-neighbor Fe-Ti and Fe-Sb forming tetrahedrons, as well as second-nearest Ti-Sb and $Fe^{4d}$-$Fe^{4c}$ forming octahedrons (Figure 2d). It is the $Fe^{4d}$-$Fe^{4c}$ interaction that introduces the *local* *gaps* and distinguishes the vacancy-containing Heuslers from the half-Heuslers. For comparison, we start by analyzing the Ti-Sb interactions, where hybridization between Ti-$e_g$ orbitals and Sb-$p$ orbitals leads to lower-energy bonding states and higher-energy antibonding states (see the inset of Figure 2b). The bonding (antibonding) states are predominantly derived from Sb-$p$ (Ti-$e_g$) orbitals because the Sb-$p$ is considerably lower in energy. By contrast, the $Fe^{4c}$-site $d$-orbitals and $Fe^{4d}$-site $d$-orbitals, which energetically degenerate before hybridization, contribute almost equally to the bonding and antibonding states. Moreover, the $e_g$-$e_g$ orbital interaction is more pronounced than the $t_{2g}$-$t_{2g}$ interaction due to the directional characteristics of the $e_g$ orbitals, resulting in a *local gap* between two groups of antibonding states (see the inset of Figure 2b). Consequently, Fe-$e_g$ orbitals are partially occupied, contrary to the empty scenario from the nominal valence [20]; Fe-$t_{2g}$ orbitals are not fully occupied, as they also hybridize with Ti-3$d$ orbitals at similar energy levels. We conclude that the fundamental bandgap mechanism in $Fe_{1.5}TiSb$ mirrors that in integer-stoichiometric full-Heuslers [3]. However, the environmental differences at the $Fe^{4c}$ and $Fe^{4d}$ sites, i.e., $e_g$ orbitals at the 4$d$ sites face more neighboring repulsions than those at the 4$c$ sites (Figure 1h), reducing 4$d$-site $e_g$ occupations to some extent. This is evidenced by the reduced electron counts in the $Fe^{4d}$-site $dz^2$ (0.98 electrons) and $dx^2$-$y^2$ (0.95 electrons) orbitals compared to those in the $Fe^{4c}$-site $dz^2$ (1.30 electrons) and $dx^2$-$y^2$ (1.27 electrons) orbitals.

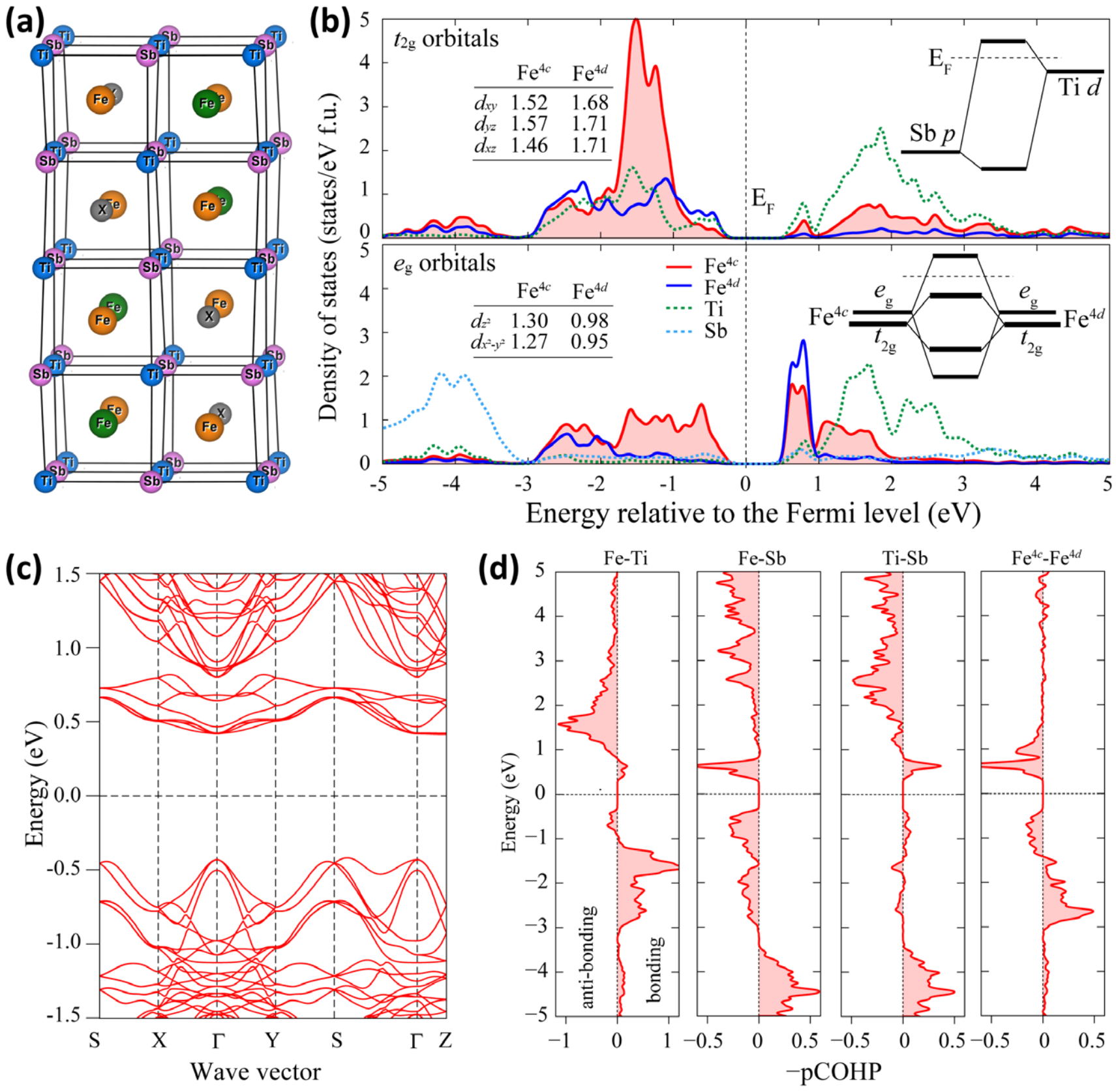

**Figure 2. Bandgap formation in $Fe_{1.5}TiSb$.** (a) Structural model. (b) Electronic density-of-states for one $Fe_{1.5}TiSb$ formula, containing one $Fe^{4c}$ atom and a half $Fe^{4d}$ atom. The electron numbers in the inset are in units of electrons per orbital. The insets also show the dominant orbital interactions, inspired by Ref. [3]. (c) Band structure. (d) Bonding and antibonding interactions characterized by negative values of *projected Crystal Orbital Hamilton Population* (–pCOHP). Four pair interactions are considered: Fe-$d$ with Ti-$d$, Fe-$d$ with Sb-$p$, Ti-$d$ with Sb-$p$, and $Fe^{4c}$-site $d$ with $Fe^{4d}$-site $d$ orbitals.

### 3. Bandgap engineering and Electron number rules

After understanding the bandgap mechanisms, we demonstrate how to transform a metallic XYZ into a semiconducting XYZ+$X'_{\beta}$, which involves selecting the X' element and determining the filling coefficient $\beta$. Figure 3a starts with half-Heuslers with $N_t$ = 18, which are semiconductors following the $s^2p^6d^{10}$ gapping mechanism. Most other

XYZ systems are metals with either deficient electrons ($N_t$ = 14 ~ 17) or excess electrons ($N_t$ = 19 ~ 20). In the electron-deficient FeTiSb, for example, we apply the $s^2p^6d^{10}+\beta\times d^6$ mechanism by introducing a homo-element of $Fe_\beta$ to the 4$d$ sites. The process of determining the $\beta$ coefficient in the targeted FeTiSb+$Fe_\beta$ is shown in the inset of Figure 3b: The FeTiSb framework accepts one electron from the $Fe_\beta$ atoms, achieving the 18-electron configuration. Since the $Fe_\beta$ atoms originally possess 8$\beta$ electrons, the remaining 8$\beta$ − 1 electrons should fully fill the 4$d$-site $t_{2g}$ orbitals with six electrons, leading to $8\beta - 1 = 6\beta$ and $\beta = 0.5$.

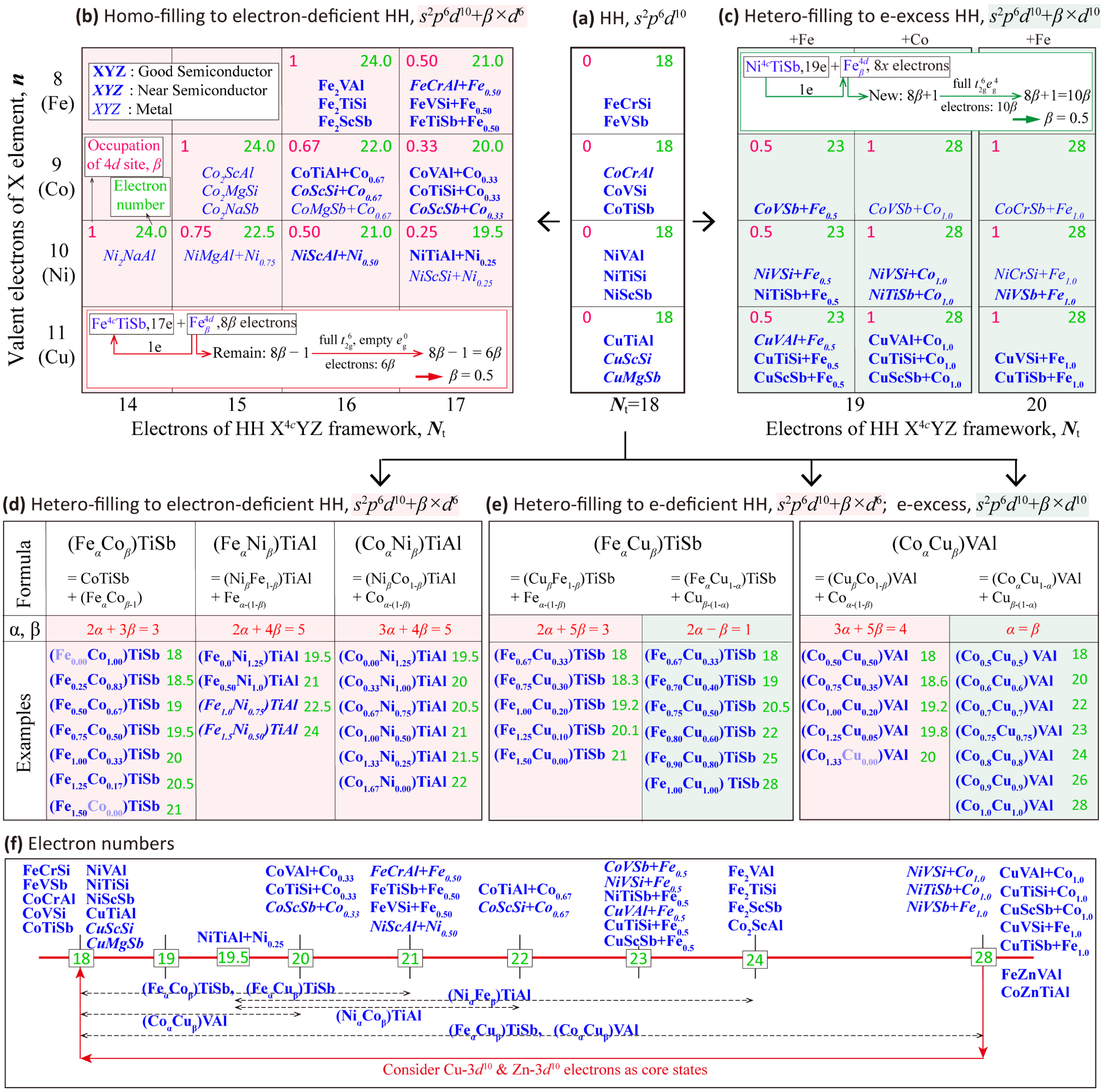

**Figure 3. Half Heuslers and their transformation into semiconductors.** (a) Half-Heuslers with 18 electrons. (b) Electron-deficient half-Heusler XYZ employs the $s^2p^6d^{10}+\beta\times d^6$ bandgap mechanism by filling the 4$d$ sites with a homo-element X. (c) Electron-excess half-Heusler XYZ utilizes the $s^2p^6d^{10}+\beta\times d^{10}$ mechanism via filling the 4$d$ sites with a hetero-element X'. (d) Hetero-filling of the electron-deficient half-Heuslers. In ($Fe_\alpha Co_\beta$)TiSb, ($Ni_\alpha Fe_\beta$)TiAl, and ($Ni_\alpha Co_\beta$)TiAl, the compositions change continuously. (e) Hetero-filling Cu-based systems implement both the $s^2p^6d^{10}+\beta\times d^6$ and $s^2p^6d^{10}+\beta\times d^{10}$ mechanisms. (f) Heusler semiconductors are categorized by their valence electron counts. Material formulas are styled in **bold** for those with a well-recognized bandgap in the simulation, in ***bold italics*** for near-semiconducting materials that represent either reality or theoretical misestimations, and in *thin italics* for those identified as metallic.

Engineering on NiTiSb relies on the $s^2p^6d^{10}$+$\beta\times d^{10}$ mechanism since NiTiSb already contains one excess electron than 18. A hetero-element with lower valence, for example Fe, is introduced to the 4*d* sites to accommodate the excess electron. As shown in Figure 3c (the inset), the NiTiSb framework donates the excess electron to the $Fe_\beta$ atoms, achieving the 18-electron configuration. The $Fe_\beta$ atoms, now possessing $8\beta + 1$ electrons, can create a *local gap* if all the *d*-orbitals are filled to a $d^{10}$ configuration, resulting in $8\beta + 1 = 10\beta$ and $\beta = 0.5$.

$Fe_\alpha$TiSb ($0 < \alpha < 1$) represents a more generic formulation where the Fe atoms are insufficient to reach a half-Heusler concentration. It can be transformed into semiconductors when filling with Co atoms through the $s^2p^6d^{10}$+$\beta\times d^6$ mechanism, resulting in $Fe_\alpha$TiSb+$Co_\beta$ or equivalently ($Fe_\alpha Co_\beta$)TiSb (Figure 3d). For conceptual simplicity, we assume the XYZ framework is CoTiSb, leaving the other atoms (i.e., $Fe_\alpha$ and $Co_{\beta-1}$) to occupy the 4*d* sites. This arrangement creates 4*d* *local gaps* under the condition that $8\alpha + 9(\beta - 1) = 6(\alpha + \beta - 1)$, of which a pictorial explanation is provided in **Section 2** of the Supplementary Materials. The resulting relationship, $2\alpha + 3\beta = 3$, indicates that the compositions are continuously variable. Figure 3e also demonstrates an example of introducing the Cu element to $Fe_\alpha$TiSb, revealing an interesting phenomenon in ($Fe_\alpha Cu_\beta$)TiSb: both gapping mechanisms are applicable and lead to two stoichiometry relations. For the $s^2p^6d^{10}$+$\beta\times d^6$ mechanism producing $2\alpha + 5\beta = 3$, all the Cu atoms enter 4*c* sites, while all the inserted atoms at 4*d* sites are entirely of Fe atoms. By contrast, all the Fe atoms enter 4*c* sites for the $s^2p^6d^{10}$+$\beta\times d^{10}$ mechanism, which gives $2\alpha - \beta = 1$. Such arrangements are governed by site occupation rules, to be discussed in **Section 5**.

The bandgap mechanisms fundamentally dictate the electron numbers, which are the necessary conditions for the development of bandgaps. However, the predicted systems might face practical constraints, necessitating material-specific investigations. We perform analyses using density-functional theory, constructing structural models based on the criteria outlined in **Section 5**. According to the simulations, semiconductors (or *near-semiconductors* due to the theoretical problem of bandgap underestimation; also see **Section 5** of the Supplementary Materials) are particularly abundant in systems containing Fe, Co, Ni, and Cu (see Figures 3 and 4; also refer to **Section 10** of the Supplementary Materials for the complete results).

The predicted semiconductors can be organized according to their valence electron numbers, as shown in Figure 3f. Besides the well-known 18 electrons in THH and 24 in TFH, TOSH semiconductors exhibit a range of discrete electron counts including 19, 19.5, 20, 21, 22, 23, and 28. Furthermore, QOSH semiconductors continuously span the range from 18 to 28 electrons. This finding demonstrates that the vacancy-filling strategy significantly expands the family of semiconducting Heuslers. The 28-electron systems, classified as XX'YZ QFH, warrant attention. The density-of-states analysis indicates that Cu-3*d* orbitals are consistently fully occupied and positioned deep away from the band edge (Figure 4). For example, the Cu-3*d* states in CuTiSb+$Fe_{1.0}$ are notably deeper than the higher-energy Fe-3*d* states. In CuVAl+$Co_{1.0}$, there is a slight overlap between Cu-3*d* and Co-3*d* orbitals, though it is minimal. From a theoretical standpoint, it is more consistent to include Cu-3*d* electrons in the valence electron count. By contrast, Zn-3*d* electrons are positioned even deeper than Cu-3*d* electrons (see **Section 6** of the Supplementary Materials), which justifies the flexibility of excluding them from the valence electron count. These analyses highlight the connection between 28-electron and 18-electron Heuslers materials.

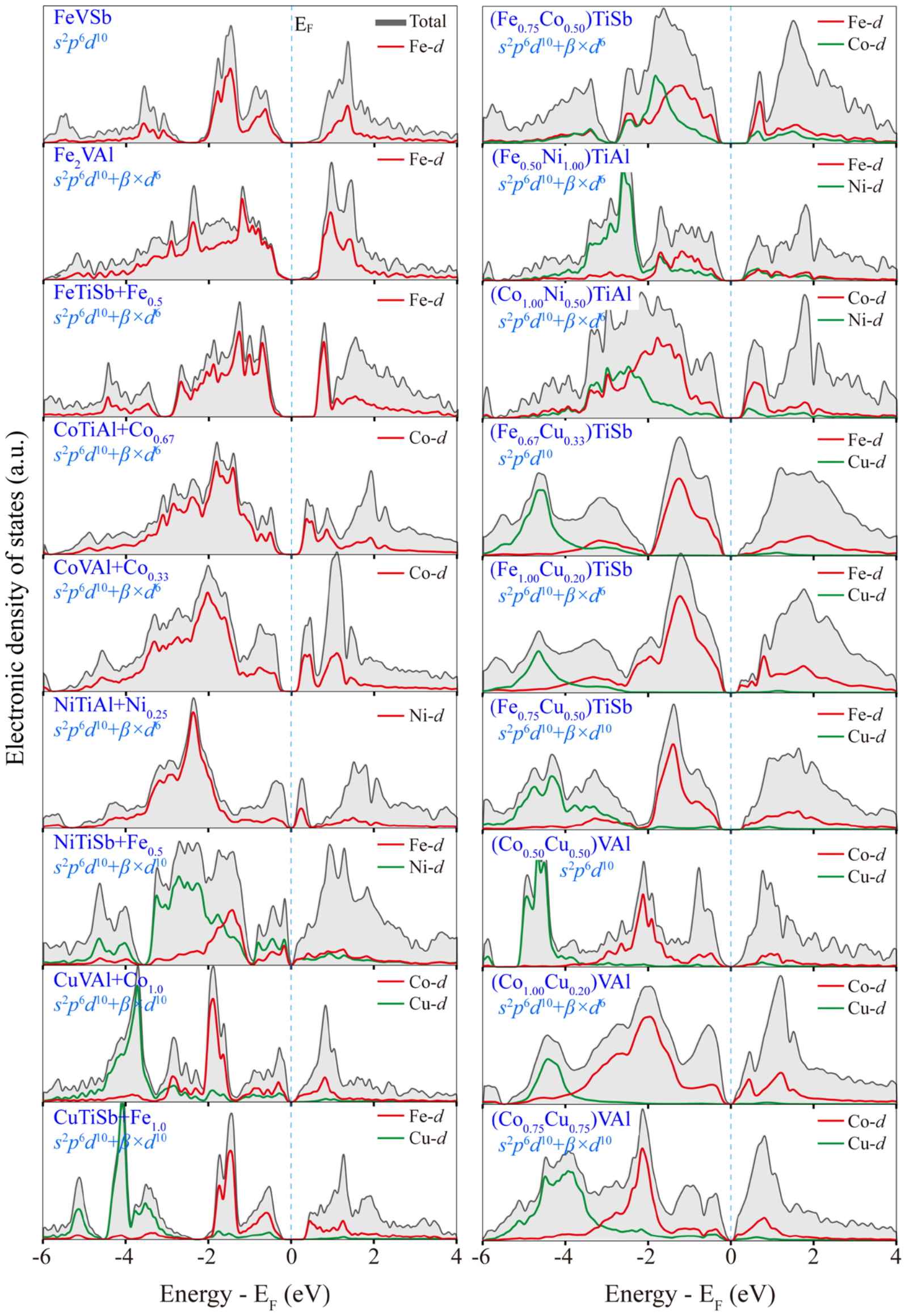

**Figure 4. Electronic density-of-states of selected Heusler semiconductors**

## 4. Comprehensive landscape

Exploring the most generic formula for Heuslers, i.e., $X_\alpha X'_\beta YZ$ for QOSH, unveils a comprehensive landscape of semiconducting Heuslers, as depicted in Figure 5. This figure incorporates all the bandgap mechanisms—$s^2p^6d^{10}$, $s^2p^6d^{10}+\beta\times d^6$, and $s^2p^6d^{10}+\beta\times d^{10}$—and covers both integer-stoichiometric and off-stoichiometric Heuslers.

We discuss several intriguing trends. First, semiconductors derived from the $s^2p^6d^{10}+\beta\times d^6$ mechanism are more numerous than those from the $s^2p^6d^{10}+\beta\times d^{10}$ mechanism. In the former situation, $e_g$-$e_g$ interaction [3] is typically significant (especially for larger ions), which ensures the ability to open a bandgap. In contrast, the latter mechanism initiates a bandgap atop the $d^{10}$ orbitals. These systems often have densely populated electrons where mutual repulsion tends to minimize and eventually close the bandgap. Second, the landscape presents many new off-stoichiometric half- and full-Heuslers, which are particularly interesting and simpler regarding theoretical understanding and experimental realization. For QHH half-Heuslers, semiconductors or near-semiconductors include $Fe_{0.5}Ni_{0.5}TiSb$, $Fe_{0.33}Cu_{0.67}VAl$, $Fe_{0.67}Cu_{0.33}TiSb$, $Co_{0.5}Cu_{0.5}VAl$, $Fe_{0.25}Zn_{0.75}TiAl$, $Fe_{0.5}Zn_{0.5}VAl$, $Fe_{0.75}Zn_{0.25}TiSb$, $Co_{0.33}Zn_{0.67}TiAl$, and $Co_{0.67}Zn_{0.33}VAl$. QFH full-Heuslers contains $Fe_{1.5}Ni_{0.5}TiAl$, $Fe_{1.33}Cu_{0.67}ScAl$, $Fe_{1.67}Cu_{0.33}TiAl$, $Fe_{1.75}Zn_{0.25}TiAl$, $Fe_{0.5}Ni_{1.5}TiAl$, $Fe_{0.33}Cu_{1.67}TiAl$, $Fe_{0.67}Cu_{1.33}VAl$, $Co_{0.5}Cu_{0.5}TiAl$, $Fe_{0.75}Zn_{1.25}TiAl$, FeCuTiSb, FeZnVAl, CoCuVAl, and CoZnTiAl. The richness of Al-based semiconductors may be related to the particular energy of Al-3$p$ orbitals [27]. More candidates are anticipated when substituting the 4$a$ and 4$b$ atoms [26]. Third, when using Ni as the filling element under the $s^2p^6d^{10}+\beta\times d^{10}$ mechanism, there are many systems without constraints on the filling fraction $\beta$ [28]. For instance, $Fe_{\alpha}TiSb+Ni_{\beta}$ conforms to the relationship of $2\alpha = 1$, indicating that $Fe_{0.5}TiSb$ can accommodate Ni atoms in amounts ranging from $\beta = 0.5$ to 1.5 (due to the constrain of $1 \le \alpha + \beta \le 2$, instead of electron number consideration). While the full-Heusler $Fe_{0.5}Ni_{1.5}TiSb$ exhibits near-semiconductor behavior, systems with lower Ni concentrations (e.g., from $Fe_{0.5}Ni_{0.5}TiSb$ to $Fe_{0.5}Ni_{1.0}TiSb$) are good semiconductors (see **Section 9** of the Supplementary Materials). A similar phenomenon has been observed in $FeNbSb+Cr_{\beta}$ systems [24], where arbitrary additions of Cr atoms (having six valence electrons) consistently result in semiconductors following the $s^2p^6d^{10}+\beta\times d^6$ bandgap mechanism.

| YZ electrons | Mechanisms of $s^2p^6d^{10}$ and $s^2p^6d^{10}+\beta\times d^6$ | | | | Mechanism of $s^2p^6d^{10}+\beta\times d^{10}$ | | | |
|---|---|---|---|---|---|---|---|---|
| | $+ Co_{\beta}$ | $+ Ni_{\beta}$ | $+ Cu_{\beta}$ | $+ Zn_{\beta}$ | $+ Co_{\beta}$ | $+ Ni_{\beta}$ | $+ Cu_{\beta}$ | $+ Zn_{\beta}$ |
| $N_{YZ} = 6$<br>e.g., $Fe_{\alpha}$ ScAl | $2\alpha + 3\beta = 6$<br>$Co_2ScAl$ | $2\alpha + 4\beta = 6$<br>$FeNiScAl$ | $2\alpha + 5\beta = 6$<br>$Fe_{1.33}Cu_{0.67}ScAl$ | $2\alpha + 6\beta = 6$<br>$ZnScAl$<br>$Fe_{1.5}Zn_{0.5}ScAl$ | $2\alpha + \beta = -2$ | $2\alpha = -2$ | $2\alpha - \beta = -2$<br>$Cu_2ScAl$ | $2\alpha - 2\beta = -2$<br>$Fe_{0.5}Zn_{1.5}ScAl$ |
| $N_{YZ} = 7$<br>e.g., $Fe_{\alpha}$ TiAl | $2\alpha + 3\beta = 5$<br>$FeCoTiAl$ | $2\alpha + 4\beta = 5$<br>$Fe_{1.5}Ni_{0.5}TiAl$ | $2\alpha + 5\beta = 5$<br>$CuTiAl$<br>$Fe_{1.67}Cu_{0.33}TiAl$ | $2\alpha + 6\beta = 5$<br>$Fe_{0.25}Zn_{0.75}TiAl$<br>$Fe_{1.75}Zn_{0.25}TiAl$ | $2\alpha + \beta = -1$ | $2\alpha = -1$ | $2\alpha - \beta = -1$<br>$Fe_{0.33}Cu_{1.67}TiAl$ | $2\alpha - 2\beta = -1$<br>$Fe_{0.75}Zn_{1.25}TiAl$ |
| $N_{YZ} = 8$<br>e.g., $Fe_{\alpha}$ VAl | $2\alpha + 3\beta = 4$<br>$Fe_2VAl$ | $2\alpha + 4\beta = 4$<br>$NiVAl$<br>$Fe_2VAl$ | $2\alpha + 5\beta = 4$<br>$Fe_{0.33}Cu_{0.67}VAl$<br>$Fe_2VAl$ | $2\alpha + 6\beta = 4$<br>$Fe_{0.5}Zn_{0.5}VAl$<br>$Fe_2VAl$ | $2\alpha + \beta = 0$ | $2\alpha = 0$<br>$Ni_2VAl$ | $2\alpha - \beta = 0$<br>$Fe_{0.67}Cu_{1.33}VAl$ | $2\alpha - 2\beta = 0$<br>$FeZnVAl$ |
| $N_{YZ} = 9$<br>e.g., $Fe_{\alpha}$ TiSb | $2\alpha + 3\beta = 3$<br>$CoTiSb$ | $2\alpha + 4\beta = 3$<br>$Fe_{0.5}Ni_{0.5}TiSb$ | $2\alpha + 5\beta = 3$<br>$Fe_{0.67}Cu_{0.33}TiSb$ | $2\alpha + 6\beta = 3$<br>$Fe_{0.75}Zn_{0.25}TiSb$ | $2\alpha + \beta = 1$ | $2\alpha = 1$<br>$Fe_{0.5}Ni_{1.5}TiSb$ | $2\alpha - \beta = 1$<br>$FeCuTiSb$ | $2\alpha - 2\beta = 1$<br>$Fe_{1.25}Zn_{0.75}TiSb$ |
| $N_{YZ} = 10$<br>e.g., $Fe_{\alpha}$ VSb | $2\alpha + 3\beta = 2$<br>$FeVSb$ | $2\alpha + 4\beta = 2$<br>$FeVSb$ | $2\alpha + 5\beta = 2$<br>$FeVSb$ | $2\alpha + 6\beta = 2$<br>$FeVSb$ | $2\alpha + \beta = 2$<br>$Co_2VSb$ | $2\alpha = 2$<br>$FeNiVSb$ | $2\alpha - \beta = 2$<br>$Fe_{1.33}Cu_{0.67}VSb$ | $2\alpha - 2\beta = 2$<br>$Fe_{1.5}Zn_{0.5}VSb$ |
| $N_{YZ} = 6$<br>e.g., $Co_{\alpha}$ ScAl | $Co_2ScAl$ | $3\alpha + 4\beta = 6$<br>$Co_2ScAl$ | $3\alpha + 5\beta = 6$<br>$Co_2ScAl$ | $3\alpha + 6\beta = 6$<br>$ZnScAl$<br>$Co_2ScAl$ | | $\alpha = -2$ | $\alpha - \beta = -2$<br>$Cu_2ScAl$ | $\alpha - 2\beta = -2$<br>$Co_{0.67}Zn_{1.33}ScAl$ |
| $N_{YZ} = 7$<br>e.g., $Co_{\alpha}$ TiAl | $CoTiAl + Co_{2/3}$ | $3\alpha + 4\beta = 5$ | $3\alpha + 5\beta = 5$<br>$CuTiAl$ | $3\alpha + 6\beta = 5$<br>$Co_{0.33}Zn_{0.67}TiAl$ | | $\alpha = -1$ | $\alpha - \beta = -1$<br>$Co_{0.5}Cu_{1.5}TiAl$ | $\alpha - 2\beta = -1$<br>$CoZnTiAl$ |
| $N_{YZ} = 8$<br>e.g., $Co_{\alpha}$ VAl | $CoVAl + Co_{1/3}$ | $3\alpha + 4\beta = 4$<br>$NiVAl$ | $3\alpha + 5\beta = 4$<br>$Co_{0.5}Cu_{0.5}VAl$ | $3\alpha + 6\beta = 4$<br>$Co_{0.67}Zn_{0.33}VAl$ | XYZ<br>Good Semiconductor | $\alpha = 0$<br>$Ni_2VAl$ | $\alpha - \beta = 0$<br>$CoCuVAl$ | $\alpha - 2\beta = 0$<br>$Co_{1.33}Zn_{0.67}VAl$ |
| $N_{YZ} = 9$<br>e.g., $Co_{\alpha}$ TiSb | $CoTiSb$ | $3\alpha + 4\beta = 3$<br>$CoTiSb$ | $3\alpha + 5\beta = 3$<br>$CoTiSb$ | $3\alpha + 6\beta = 3$<br>$CoTiSb$ | XYZ<br>Near Semiconductor | $\alpha = 1$<br>$CoNiTiSb$ | $\alpha - \beta = 1$<br>$Co_{1.5}Cu_{0.5}TiSb$ | $\alpha - 2\beta = 1$<br>$Co_{1.67}Zn_{0.33}TiSb$ |
| $N_{YZ} = 10$<br>e.g., $Co_{\alpha}$ VSb | | $3\alpha + 4\beta = 2$ | $3\alpha + 5\beta = 2$ | $3\alpha + 6\beta = 2$ | XYZ<br>Metal | $\alpha = 2$<br>$Co_2VSb$ | $\alpha - \beta = 2$<br>$Co_2VSb$ | $\alpha - 2\beta = 2$<br>$Co_2VSb$ |

Figure 5. **Relationship of the compositions $\alpha$ and $\beta$ in the generic formulation $X_{\alpha}X'_{\beta}YZ$.** The $\alpha$-$\beta$ relationship is subject to two constraints: ensuring the electron count is suitable for semiconductors (the relationship is indicated in red font) and maintaining an atomic concentration such that $1 \le \alpha + \beta \le 2$. The endpoints of half-Heuslers ($\alpha + \beta = 1$) and full-Heuslers ($\alpha + \beta = 2$) are shown in blue fonts. The hashed areas denote that $\alpha$ and $\beta$ cannot fulfill a logical relation. The shaded elliptical areas highlight regions more likely to achieve good semiconductors according to density-functional-theory predictions. The first column provides examples of Fe- and Co-based systems, which can be generalized to other elements.

## 5. Atomic occupation rules and Emerging weak metallicity

Due to off-stoichiometry, occupation disordering may occur and strongly influence the material properties. Understanding the fundamental principles of atomic occupations is crucial for constructing useful structural models. In this section, we derive several structural rules and explore the consequences of their violations.

When discussing bandgap mechanisms, we have posited that the 4$c$ sites are preferentially occupied prior to filling the 4$d$ sites. This ordering principle, defined as the ***4c's full occupation rule***, is driven by the strong desire to minimize Coulombic repulsion between the 4$c$ and 4$d$ sites. Following this, the 4$d$ sites are then partially filled with atoms, resulting in numerous possible occupation patterns. An ordered distribution of atoms across the 4$d$ sites, referred to as the ***4d's ordered distribution rule***, is found energetically most favorable, as it further diminishes the 4$c$-4$d$ repulsion. Moreover, if the 4$c$ sites contain two different types of atoms, they are likely to be uniformly mixed.

Quaternary $X_\alpha X'_\beta YZ$ is subject to an additional rule in arranging X and X' ions. For example, $(Fe_\alpha Cu_\beta)TiSb$ can utilize both $s^2p^6d^{10}+\beta\times d^6$ and $s^2p^6d^{10}+\beta\times d^{10}$ mechanisms to form two types of semiconductors, each associated with distinct 4*c*-4*d* occupation patterns for the Fe and Cu atoms. This behavior is governed by the ***4d's size selection rule***, which is motivated by the objective of harvesting energy from bandgap formation. To open a bandgap under the $s^2p^6d^{10}+\beta\times d^6$ mechanism, the 4*d* sites preferentially host larger ions to enhance 4*c*-4*d* interactions, thereby increasing the $e_g$-$e_g$ orbital interaction (Figure 6a). Conversely, the $s^2p^6d^{10}+\beta\times d^{10}$ mechanism favors smaller ions at the 4*d* sites, keeping the $d^{10}$ orbitals more compact and creating a gap from higher electronic states. Applying these principles to $(Fe_\alpha Cu_\beta)TiSb$ [23], the $s^2p^6d^{10}+\beta\times d^6$ semiconductors require 4*d* sites occupied by the larger Fe, while the $s^2p^6d^{10}+\beta\times d^{10}$ semiconductors have their 4*d* sites filled with the smaller Cu ions.

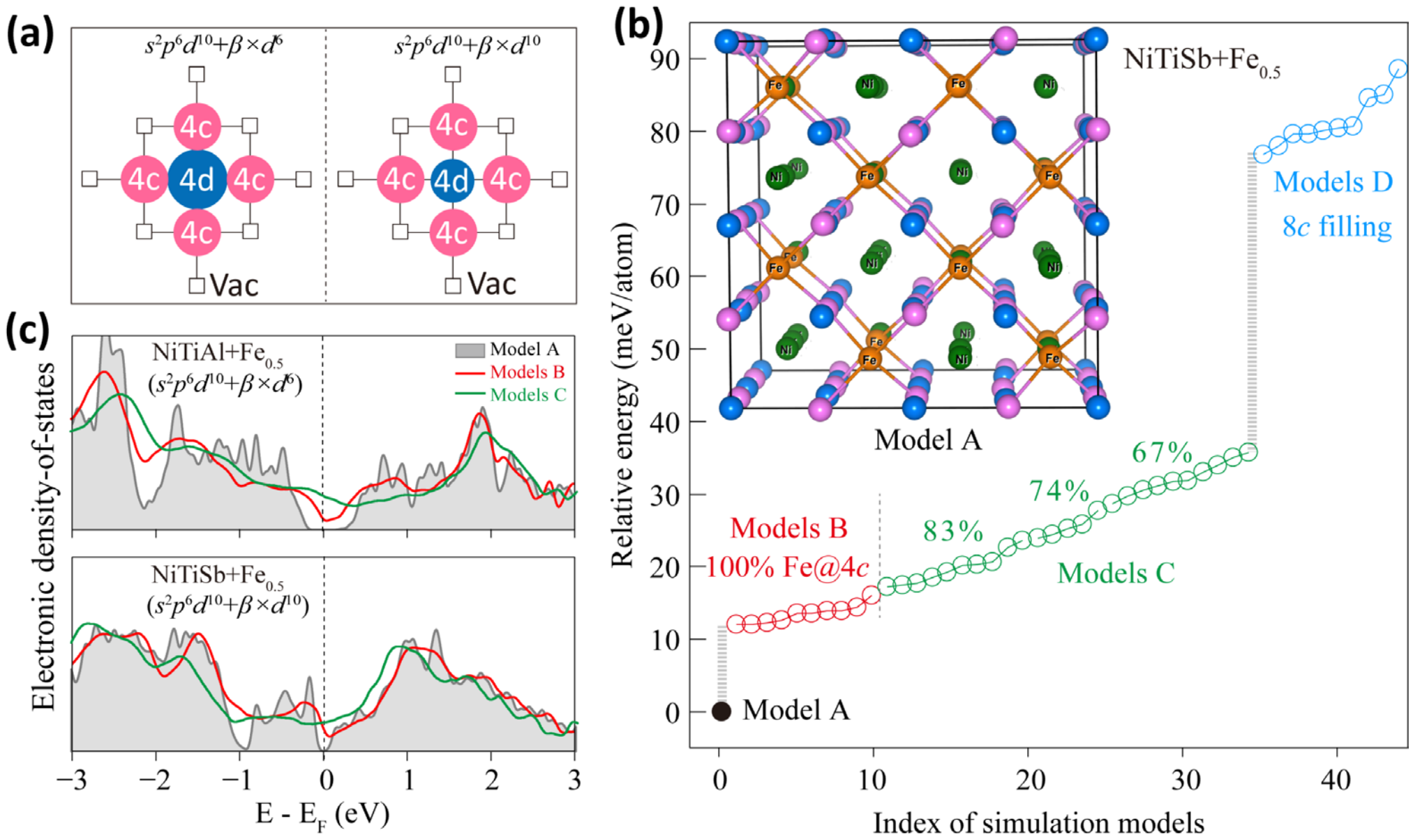

**Figure 6. Occupation rules and consequences of deviation.** (a) Sketch of the *4d's size selection rule* designed to facilitate the semiconducting phase of quaternary $X_\alpha X'_\beta YZ$. (b) Energy penalties associated with deviating from the occupation rules. The inset shows an ideal ground state, Model A, for NiTiSb+$Fe_{0.5}$. The structure is a 2×2×2 supercell containing 112 atoms in total, including 32 Ni and 16 Fe atoms; while 16 Ni and 16 Fe atoms ordered fill all the 4*c* sites, the remaining 16 Ni atoms evenly occupy half of the 4d sites. Model B violates the *4d's ordered distribution rule*, Model C breaches the *4d's size selection rule*, and Model D also flouts the *4c's full occupation rule*. Models B through D utilize larger supercells with 378 atoms to better account for occupation disorder. (c) Density-of-states for NiTiAl+$Fe_{0.5}$ and NiTiSb+$Fe_{0.5}$. For Models B and C, each result represents an average taken from many supercell structures.

While electron number rules can be rigorously enforced through composition control, adherence to occupation rules may be more challenging in practice. In this context, we explore the consequences of such violations in two Ni-based systems (Figure 6): NiTiAl+$Fe_{0.5}$ and NiTiSb+$Fe_{0.5}$, which follows the $s^2p^6d^{10}+\beta\times d^6$ and $s^2p^6d^{10}+\beta\times d^{10}$ mechanisms, respectively. To apply the atomic occupation rules, supercell models with 2×2×2 unitcell size are constructed. The first two occupation rules can be readily implemented in this supercell, but applying the ***4d's size selection rule*** presents certain complexities. In NiTiAl+$Fe_{0.5}$, all the larger Fe atoms are strategically placed in the 4*d* sites; in NiTiSb+$Fe_{0.5}$, all Fe atoms are placed in the 4*c* sites. Model A (see the inset of Figure 6b), which rigorously adheres to all the rules, exhibits semiconducting properties with bandgaps of 0.43 eV and 0.11 eV for NiTiAl+$Fe_{0.5}$ and NiTiSb+$Fe_{0.5}$, respectively (Figure 6c).

Any deviation from the ideal occupations destabilizes the structure, as demonstrated by the penalizing energies in NiTiSb+$Fe_{0.5}$ (Figure 6b). Violating the ***4d's ordered distribution rule***, likely under finite temperatures, results in the partial and random occupation of 4*d* sites by inadequate Ni atoms. This randomness, considered in larger supercells (Models B), leads to increased energies of at least 11 meV/atom and the emergence of weak metallicity (Figure 6c). Violations of the ***4d's size selection rule*** occur as Fe atoms gradually fill 4*d* sites (Models C), progressing to a completely random mix of Fe and Ni (67% Fe remaining in 4*c* sites), which increases energy and enhances metallic behavior. Finally, the ***4c's full*** ***occupation rule*** is violated when Fe and Ni mix randomly in all interstitial sites, transforming the 4*c* and 4*d* Wyckoff positions into a single 8c symmetry (Models D). These structures exhibit significantly higher energies and are energetically discontinuous from the earlier models.

## 6. Experimental verification

Experimentally, we have successfully synthesized polycrystalline samples of several systems. The X-ray diffraction (XRD) patterns for these samples are presented in Figure 7a, and the corresponding Rietveld refinements can be found in Figure S7 of the Supplementary Materials. Theoretical analysis reveals that quaternary half-Heusler $(Fe_{0.67}Cu_{0.33})TiSb$ [23] adheres to the $s^2p^6d^{10}$ mechanism. The QOSH $(FeCu_{0.2})TiSb$ [25] relies on the $s^2p^6d^{10}+\beta\times d^6$ bandgap mechanism, while (FeCu)TiSb and $(Fe_{0.75}Cu_{0.50})TiSb$ [23] utilize the $s^2p^6d^{10}+\beta\times d^{10}$ mechanism. The fundamental characteristic of semiconductors is clearly evident from the presence of optical bandgaps, as shown in the inset of Figure 7a. Detailed electrical transport properties (**Section 4.3** of the Supplementary Materials) reveal some characteristics typical of degenerate semiconductors, which correlate with the atomic occupations discussed in **Section 5**.

Independent research [26] has investigated the thermoelectric properties of quinary $(Co_{0.85}Cu_{0.65})Zr(Sn_{1-y}Sb_y)$ with $0 \leq y \leq 0.25$, which are utilized to test our theoretical predictions. The X-ray diffraction patterns (Figure 7b) indicate that the samples with $0 \leq y \leq 0.2$ are high-quality crystals, whereas the sample with $y = 0.25$ displays a secondary phase. $(Co_{0.85}Cu_{0.65})Zr(Sn_{0.8}Sb_{0.2})$ exhibits the lowest electrical conductivity and the highest Seebeck coefficient (~$0.2\times10^5$ S/m and ~50 μV/K at 300 K), suggesting that this sample closely approximates semiconductor behavior. Theoretically, we first construct superstructures and then calculate the electronic properties (see **Section 4.1** of the Supplementary Materials). Our theoretical simulations (Figure 7c) show that the samples with $y < 0.20$ behave as *p*-type metals, while the sample with $y = 0.20$ acts as an intrinsic semiconductor. $(Co_{0.85}Cu_{0.65})Zr(Sn_{0.8}Sb_{0.2})$ adheres to the $s^2p^6d^{10}+\beta\times d^{10}$ bandgap mechanism since the nominal orbital configurations are Co-$d^{10}$, Cu-$d^{10}$, Zr-$d^0$, Sn-$s^2p^6$, and Sb-$s^2p^6$. An even higher electron concentration at $y = 0.25$ results in *n*-type behavior, which explains the experimental difficulty in realizing a single phase.

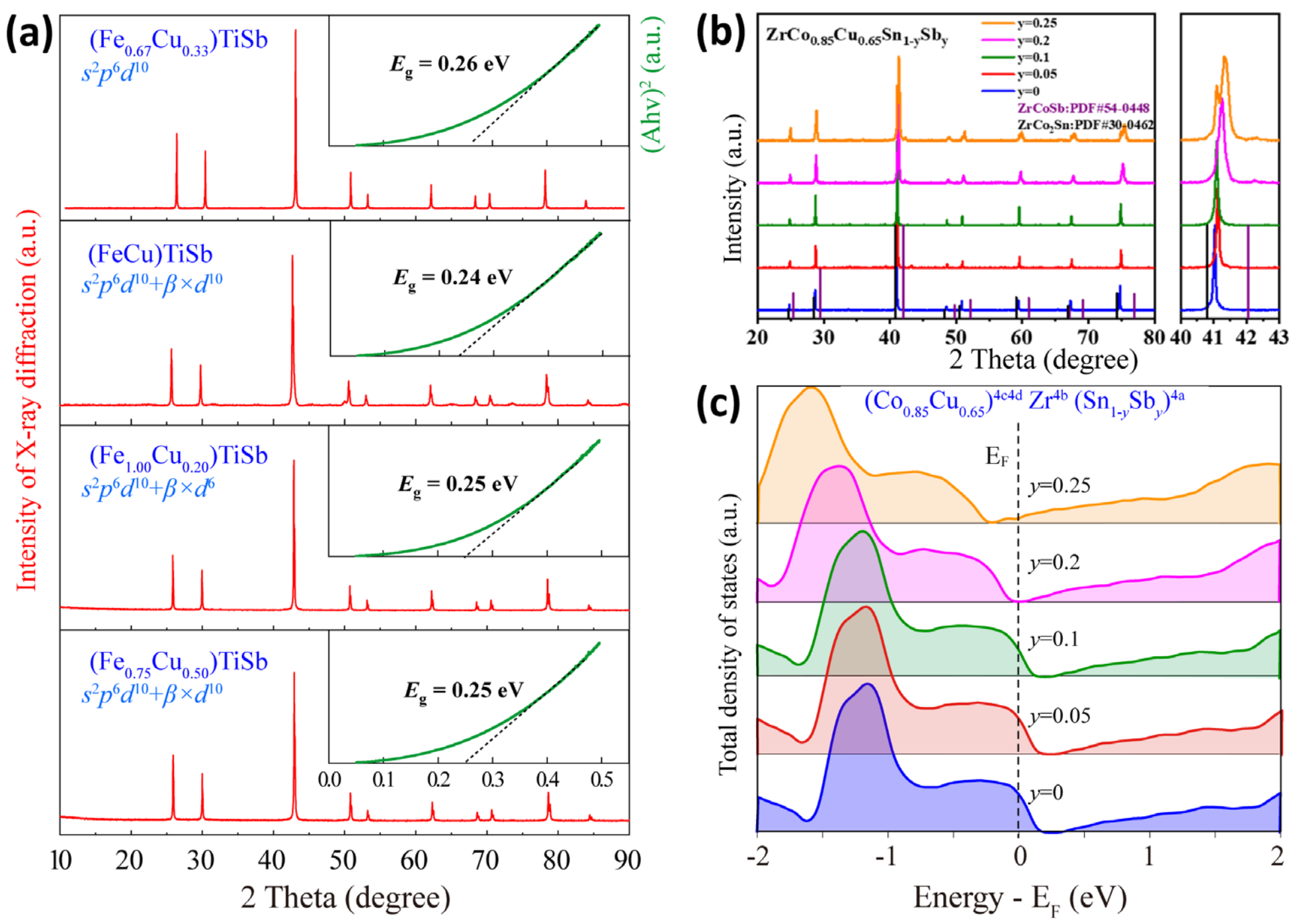

**Figure 7. Characterization of selected Heusler semiconductors.** (a) X-ray diffraction (XRD) patterns of the $(Fe_\alpha Cu_\beta)TiSb$ samples, with insets showing the determination of optical bandgaps. (b) XRD of quinary $(Co_{0.85}Cu_{0.65})Zr(Sn_{1-y}Sb_y)$ with $0 \leq y \leq 0.25$. Note the double-peak structures for the $y = 0.25$ sample, indicating the presence of a secondary phase. Reproduced from Ref. [26] with permission. (c) Evolution of density-of-states for the $(Co_{0.85}Cu_{0.65})Zr(Sn_{1-y}Sb_y)$ materials.

## 7. Discussion and Perspective

### 7.1. Correlating the theoretical structures with practical experiments

Given the complexities inherent in structural models, an important question arises: How effectively can the predicted material attributes be implemented in practical experiments? As an illustrative example, we reconsider $Fe_{1.5}TiSb$ (see **Section 3** of the Supplementary Materials), the first off-stoichiometric semiconducting Heusler [18]. Figure S2 displays five structural models with increasing complexities. Model-I (space group $R3m$), identified using the cluster expansion approach to explore the configurational space [18], has the lowest energy among all the tested models (Table S1). Model-II (space group $I\bar{4}2d$) shows a relative higher energy of 16 meV/atom. Both Model-I and Model-II are small and exhibit both short- and long-range ordering of the 4*d* atoms, conforming to the ***4d's ordered distribution rule***. Model-III, which is larger and includes some 4*d* disorder, exhibits a higher energy of 28 meV/atom. The even larger Model-IV, which displays significant lattice distortions that more effectively accommodate 4*d* disorder, has an energy of 14 meV/atom. These models all conform to the ***4c's full occupation rule*** and are semiconductors (Figure S3). Further breaking this rule, as conducted in Model-V, leads to metallicity and a significantly higher energy of 96 meV/atom. Therefore, bandgap formation serves as a strong driving force to stabilize the crystal structures, providing evidence for the material design strategies proposed in this study.

Vibrational dynamic stability, typically assessed through phonon dispersion analyses, is critical for the off-stoichiometric Heuslers. Models I through III consistently display positive vibrational frequencies (Figure S4), confirming their dynamic stability. Model-IV, marked by significant disorder, may compromise the applicability of phonon analysis. Instead, their vibrational properties are explored using molecular dynamics simulations. As shown in Figure S5, all atoms well retain their positions without intersite atomic transfer, indicating phase stability. To summarize, the observation that off-stoichiometric Heuslers can accommodate meta-stable configurations highlights a distinctive characteristic of intermetallic compounds.

Thermodynamic stability, determining the favorability of a particular stoichiometry over others, is also a critical aspect of material stability. Previous theoretical investigations of the convex hull reveal that $Fe_{1.5}TiSb$ exhibits the lowest formation energy among all considered stoichiometries of Fe-Ti-Sb [18]. However, we have experimentally realized stoichiometries covering the entire range of $1.30 \leq x \leq 1.50$ for $Fe_xTiSb$ (to be published elsewhere). Additionally, in the $(Fe_\alpha Cu_\beta)TiSb$ series conforming to semiconducting stoichiometries, we demonstrated that $(Fe_{0.67}Cu_{0.33})TiSb$ has the lowest formation energy and higher stoichiometries reduce thermodynamic stability [23]. Nevertheless, (FeCu)TiSb with a stoichiometry of $\alpha = \beta = 1$ has been successfully synthesized (Figure 4b). These findings underscore the flexibility of off-stoichiometric Heuslers to accommodate a variety of stoichiometries.

### 7.2. Expanding the landscape through elemental substitution

This work focuses on 3*d*-electron transition-metal elements, including low-valent elements of Sc, Ti, V, and Cr, and high-valent elements of Fe, Co, Ni, Cu, and Zn. The element Mn is excluded from this study for two primary reasons: firstly, Mn can occupy both the 4*c* and 4*d* tetrahedral sites and the 4*b* octahedral sites, introducing significant structural complexities, such as those seen in inverse Heuslers [4]. Secondly, Mn in the 4*b* sites typically exhibits strongly localized magnetic moments, leading to distinct magnetic and metallic behaviors.

The landscape can be readily extended to elements with 4*d*/5*d*-electrons. While the electron number rules should still apply, a few cautions are worth noting (see **Section 7** of the Supplementary Materials). First, the bandgap may be reduced or even diminished due to the spin-orbit coupling effect introduced by heavy elements. Second, differences in atomic sizes can affect the structural stability compared to their 3*d*-electron counterparts. Therefore, a case-by-case investigation is necessary to identify viable 4*d*/5*d*-electron candidates.

The off-stoichiometric Heusler family can be further expanded by simultaneously exploring variations in the anionic elements. For example, Figure 4c illustrates the mixture of Sn and Sb in the form $Sn_{1-y}Sb_y$ ($0 \leq y \leq 0.25$) [26]. The effective electron number continuously changes from 4 (at $y = 0$) to 4.25 (at $y = 0.25$), introducing additional flexibility in designing off-stoichiometric Heusler semiconductors.

### 7.3. Potential impact of magnetization

The well-established semiconductors (e.g., CoTiSb, NiTiSn, and $Fe_2VAl$) do not exhibit a magnetic degree of freedom, and magnetic Heuslers are typically metals or half-metals. This observation underscores the strong interconnections between nonmagnetic semiconductors and magnetic metals, as well as the contrasting behaviors of magnetic semiconductors. Indeed, only a few exotic magnetic semiconductors have been theoretically proposed. For example, CoVTiAl is a ferromagnetic semiconductor [27], and CrVTiAl is a fully-compensated ferrimagnetic semiconductor [29], which embody these contradictory properties through the unique characteristics of the Al-3*p* orbitals [27]. However, subsequent experiments on CrVTiAl have revealed gapless behaviors that deteriorate its semiconducting properties [30]. Overall, achieving a robust merger between magnetism and semiconductor properties in Heusler compounds remains uncommon and rare. Based on these observations, we have employed spin-unpolarized simulations. While this assumption is valid for most materials, we indeed identify its limitations in some cases. For instance, we identify FeZnVAl as a semiconductor from the nonmagnetic simulations, aligning with a recent report [31]. However, spin-polarization is strongly favored in this material, effectively transforming it into a magnetic metal. Further research is underway to address the magnetic issue.

## 8. Summary

The vacancy-filling strategy, applied to the generic formula $X_\alpha X'_\beta YZ$ ($1 \le \alpha + \beta \le 2$), serves as a practical approach for expanding the range of semiconducting systems. This approach naturally encompasses the well-known integer-stoichiometric Heuslers and facilitates the prediction of numerous off-stoichiometric variants, thereby providing a comprehensive landscape of semiconducting Heusler compounds. The coefficients $\alpha$ and $\beta$ are determined based on the *valence electron counts* that contribute to semiconductor properties, and the corresponding bandgap mechanisms are categorized into three types: $s^2p^6d^{10}$, $s^2p^6d^{10}+\beta\times d^6$, and $s^2p^6d^{10}+\beta\times d^{10}$. Materials that satisfy bandgap mechanisms, which determine the necessary electron counts for the semiconducting behavior, are also subject to practical electronic and structural constraints to develop bandgaps. For example, atomic occupation patterns are crucial in determining the material properties: ideal site occupancy typically results in good semiconducting behavior, whereas occupation disorder can lead to weak metallicity.

Despite the significant complexities associated with off-stoichiometry, there are a few simple ***electron count rules*** for designing semiconducting Heusler compounds. **(1) Integer-stoichiometric Heuslers:** Half-Heuslers, represented by the generic formula $X_\alpha X'_\beta YZ$ with $\alpha + \beta = 1$, open their bandgaps through the $s^2p^6d^{10}$ mechanism, resulting in 18 valence electrons. For full-Heuslers, the ternary systems rely on the $s^2p^6d^{10}+\beta\times d^6$ bandgap mechanism and have 24 electrons, while the quaternary systems utilize the $s^2p^6d^{10}+\beta\times d^{10}$ mechanism and have 28 electrons. **(2) Vacancy-filling off-stoichiometric Heuslers** with $1 < \alpha + \beta < 2$: The bandgap mechanisms are either $s^2p^6d^{10}+\beta\times d^6$ or $s^2p^6d^{10}+\beta\times d^{10}$. While the electron counts in ternary systems are fixed at a few discrete values, they vary continuously from 18 to 28 in quaternary systems. **(3) Good semiconductors** typically have compositions proximate to half-Heuslers. Elevated atomic concentrations and electron counts can lead to a reduction or closure of the bandgap.

Unlike conventional crystalline materials with well-defined atomic positions, off-stoichiometric Heuslers can accommodate meta-stable atomic configurations, leading to flexibility in constructing their geometries. Here are the ***site occupation rules*** for deriving low-energy semiconductors. **(1)** ***4c's full occupation rule*****:** Despite the geometrical symmetry between 4*c* and 4*d* sites, atoms have a pronounced priority to first fill the 4*c* sites before entering the 4*d* sites. Violating this propensity requires overcoming a high energy barrier and leads to metallicity. **(2)** ***4d's ordered distribution rule*****:** If the 4*d* sites are partially occupied, the ordered distribution of the filling atoms is energetically favorable. This rule also applies to 4*c* sites when accommodating more than one element. Breaking this rule requires less energy than the previous rule. **(3)** ***4d's size selection rule***: Larger ions are favored at the 4*d* sites under the $s^2p^6d^{10}+\beta\times d^6$ mechanism, whereas smaller ions are preferred when the $s^2p^6d^{10}+\beta\times d^{10}$ mechanism is employed. Deviations can lead to the emergence of weak metallic behaviors.

## Acknowledgments

This work was supported by the Natural Science Foundation of China (52272226 and 11904156).

## Conflict of Interest

There are no conflicts of interest to declare.

## References

[1] Friedrich Heusler, *Über magnetische manganlegierungen.* Verh. Dtsch. Phys. Ges. 5, 219 (1903).

[2] I. Galanakis, P. H. Dederichs, N. Papanikolaou, *Origin and properties of the gap in the half-ferromagnetic Heusler alloys.* Physical Review B. 66, (2002).

[3] I. Galanakis, P. H. Dederichs, N. Papanikolaou, *Slater-Pauling behavior and origin of the half-metallicity of the full-Heusler alloys.* Physical Review B. 66, 174429 (2002).

[4] S Skaftouros, Kemal Özdoğan, E Şaşıoğlu, I Galanakis, *Generalized Slater-Pauling rule for the inverse Heusler compounds.* Physical Review B. 87, 024420 (2013).

[5] Linus Pauling, *The Nature of the Interatomic Forces in Metals.* Physical Review. 54, 899 (1938).

[6] J. C. Slater, *The Ferromagnetism of Nickel. II. Temperature Effects.* Physical Review. 49, 931 (1936).

[7] Kemal Özdoğan, E Şaşıoğlu, I Galanakis, *Slater-Pauling behavior in LiMgPdSn-type multifunctional quaternary Heusler materials: Half-metallicity, spin-gapless and magnetic semiconductors.* Journal of Applied Physics. 113, (2013).

[8] KP Bhatti, V Srivastava, DP Phelan, RD James, C Leighton, *Heusler Alloys: Properties, Growth, Applications*. 2016, Springer.

[9] Tanja Graf, Claudia Felser, Stuart S. P. Parkin, *Simple rules for the understanding of Heusler compounds.* Progress in Solid State Chemistry. 39, 1 (2011).

[10] Sheron Tavares, Kesong Yang, Marc A Meyers, *Heusler alloys: Past, properties, new alloys, and prospects.* Progress in Materials Science. 132, 101017 (2023).

[11] Shashwat Anand, Max Wood, Yi Xia, Chris Wolverton, G. Jeffrey Snyder, *Double Half-Heuslers.* Joule. 3, 1226 (2019).

[12] Zihang Liu, Shuping Guo, Yixuan Wu, Jun Mao, Qing Zhu, Hangtian Zhu, Yanzhong Pei, Jiehe Sui, Yongsheng Zhang, Zhifeng Ren, *Design of high-performance disordered half-Heusler thermoelectric materials using 18-electron rule.* Advanced Functional Materials. 29, 1905044 (2019).

[13] Airan Li, Madison K Brod, Yuechu Wang, Kejun Hu, Pengfei Nan, Shen Han, Ziheng Gao, Xinbing Zhao, Binghui Ge, Chenguang Fu, *Opening the bandgap of metallic Half-Heuslers via the introduction of d–d orbital interactions.* Advanced Science. 10, 2302086 (2023).

[14] Wolfgang G. Zeier, Shashwat Anand, Lihong Huang, Ran He, Hao Zhang, Zhifeng Ren, Chris Wolverton, G. Jeffrey Snyder, *Using the 18-Electron Rule To Understand the Nominal 19-Electron Half-Heusler NbCoSb with Nb Vacancies.* Chemistry of Materials. 29, 1210 (2017).

[15] Shashwat Anand, Kaiyang Xia, Vinay I. Hegde, Umut Aydemir, Vancho Kocevski, Tiejun Zhu, Chris Wolverton, G. Jeffrey Snyder, *A valence balanced rule for discovery of 18-electron half-Heuslers with defects.* Energy & Environmental Science. 11, 1480 (2018).

[16] Jiangang He, S. Shahab Naghavi, Vinay I. Hegde, Maximilian Amsler, Chris Wolverton, *Designing and Discovering a New Family of Semiconducting Quaternary Heusler Compounds Based on the 18-Electron Rule.* Chemistry of Materials. 30, 4978 (2018).

[17] Runan Xie, Jean-Claude Crivello, Céline Barreteau, *Screening New Quaternary Semiconductor Heusler Compounds By Machine-Learning Methods.* Chemistry of Materials. 35, 7615 (2023).

[18] N. Naghibolashrafi, S. Keshavarz, Vinay I. Hegde, A. Gupta, W. H. Butler, J. Romero, K. Munira, P. LeClair, D. Mazumdar, J. Ma, A. W. Ghosh, C. Wolverton, *Synthesis and characterization of Fe-Ti-Sb intermetallic compounds: Discovery of a new Slater-Pauling phase.* Physical Review B. 93, 104424 (2016).

[19] A. Tavassoli, A. Grytsiv, G. Rogl, V. V. Romaka, H. Michor, M. Reissner, E. Bauer, M. Zehetbauer, P. Rogl, *The half Heusler system $Ti_{1+x}Fe_{1.33-x}Sb$-TiCoSb with Sb/Sn substitution: phase relations, crystal structures and thermoelectric properties.* Dalton Trans. 47, 879 (2018).

[20] Shashwat Anand, G. Jeffrey Snyder, *Structural Understanding of the Slater–Pauling Electron Count in Defective Heusler Thermoelectric $TiFe_{1.5}Sb$ as a Valence Balanced Semiconductor.* ACS Applied Electronic Materials. 4, 3392 (2022).

[21] Luyao Wang, Zirui Dong, Shihua Tan, Jiye Zhang, Wenqing Zhang, Jun Luo, *Discovery of a Slater–Pauling Semiconductor $ZrRu_{1.5}Sb$ with Promising Thermoelectric Properties.* Advanced Functional Materials. 32, 2200438 (2022).

[22] Zirui Dong, Jun Luo, Chenyang Wang, Ying Jiang, Shihua Tan, Yubo Zhang, Yuri Grin, Zhiyang Yu, Kai Guo, Jiye Zhang, Wenqing Zhang, *Half-Heusler-like compounds with wide continuous compositions and tunable p- to n-type semiconducting thermoelectrics.* Nature Communications. 13, 35 (2022).

[23] Weimin Hu, Song Ye, Qizhu Li, Binru Zhao, Masato Hagihala, Zirui Dong, Yubo Zhang, Jiye Zhang, Shuki Torri, Jie Ma, Binghui Ge, Jun Luo, *Strategic Design and Mechanistic Understanding of Vacancy-Filling Heusler Thermoelectric Semiconductors.* Advanced Science. 11, 2407578 (2024).

[24] Jiajun Chen, Zirui Dong, Qizhu Li, Binghui Ge, Jiye Zhang, Yubo Zhang, Jun Luo, *Enhanced Thermoelectric Performance in Vacancy-Filling Heuslers due to Kondo-Like Effect.* Advanced Materials. 36, 2405858 (2024).

[25] Siyuanyang Yin, Qizhu Li, Shaoqin Wang, Jiajun Chen, Zirui Dong, Yubo Zhang, Binghui Ge, Jiye Zhang, Jun Luo, *Structure and thermoelectric properties of half-Heusler-like TiFeCuxSb alloys.* Journal of Materiomics. 10, 523 (2024).

[26] Hui Huang, *Synthesis and thermoelectric properties of defective Heusler-based compounds*. 2021, University of Chinese Academy of Sciences.

[27] I Galanakis, Kemal Özdoğan, E Şaşıoğlu, *A proposal for an alternative class of spin filter materials: Hybridization-induced high-TC ferromagnetic semiconductors CoVXAl (X= Ti, Zr, Hf).* Applied Physics Letters. 103, 142404 (2013).

[28] Yurong Ruan, Tao Feng, Ke Zhong, Bing Wen, Wenqing Zhang, *Full-shell d-orbitals of interstitial Ni and anomalous electrical transport in Ni-based half-Heusler thermoelectric semiconductors.* Materials Today Physics. 48, 101558 (2024).

[29] I Galanakis, Kemal Özdoğan, E Şaşıoglu, *High-T C fully compensated ferrimagnetic semiconductors as spin-filter materials: The case of CrVXAl (X= Ti, Zr, Hf) Heusler compounds.* Journal of Physics: Condensed Matter. 26, 086003 (2014).

[30] Gregory M Stephen, Christopher Lane, Gianina Buda, David Graf, Stanislaw Kaprzyk, Bernardo Barbiellini, Arun Bansil, Don Heiman, *Electrical and magnetic properties of thin films of the spin-filter material CrVTiAl.* Physical Review B. 99, 224207 (2019).

[31] Abhigyan Ojha, Rama Krushna Sabat, Sivaiah Bathula, *Exploring the thermoelectric performance of NiFeMnAl and ZnFeVAl as novel quaternary Heusler compounds.* Materials Science and Engineering: B. 311, 117789 (2025).